\documentclass[a4paper,11pt]{article}

\usepackage{amssymb,latexsym}

\begin{document}
\topmargin -10mm \topmargin 0pt \oddsidemargin 0mm

\renewcommand{\thefootnote}{\fnsymbol{footnote}}
\newcommand{\nn}{\nonumber\\}
\begin{titlepage}

\vspace*{10mm}
\begin{center}
{\Large \bf Generalized Second Law of Thermodynamics in Extended
Theories of Gravity}

\vspace*{20mm}

{\large M. Akbar~\footnote{Email address: ak64bar@yahoo.com}
\footnote{Email address: makbar@camp.edu.pk}}\\
\vspace{8mm} { \em Centre for Advanced Mathematics and Physics\\
National University of Sciences and Technology\\
 Peshawar Road, Rawalpindi, Pakistan}
\end{center}
\vspace{20mm} \centerline{{\bf{Abstract}}} \vspace{5mm} By employing
the general expression of temperature $T_{h}=|\kappa|/2\pi =
\frac{1}{2\pi
\tilde{r}_{A}}(1-\frac{\dot{\tilde{r}}_{A}}{2H\tilde{r}_{A}})$
associated with the apparent horizon of a FRW universe and assuming
a region of FRW universe enclosed by the apparent horizon as a
thermal system in equilibrium, we are able to show that the
generalized second law of thermodynamics holds in Gauss-Bonnet
gravity and in more general Lovelock gravity.
\end{titlepage}

\newpage
\renewcommand{\thefootnote}{\arabic{footnote}}
\setcounter{footnote}{0} \setcounter{page}{2}

\paragraph{Introduction:}
It is possible to associate the notions of temperature and entropy
with the apparent horizon of FRW universe analogous to the Hawking
temperature and entropy associated with the black hole horizon
\cite{a9,a2,a15}. The thermodynamic properties of the apparent
horizon of FRW universe has been studied in various theories of
gravity by many authors (see for examples \cite{a9, a2,jac, cai,
hct, btz,a10}). The thermodynamic extension has also been made at
the apparent horizon in the braneworld cosmology \cite{b1,b2,b3,b4}.
In case of black holes horizon, the study of horizon thermodynamic
has also been made by many authors (see for examples
\cite{ksp,Pad,PSP,az}). More recently, Cai et al \cite{hct} has
shown that by employing Clausius relation, $\delta Q = T_{h}dS_{h}$,
to the apparent horizon of a FRW universe, they are able to
reproduce the modified Friedmann equations by using quantum
corrected entropy-area relation. Since the extra higher derivative
terms in Gauss-Bonnet and Lovelock gravities can be thought of as
quantum corrections to the Einstein gravity and Friemann equations
at apparent horizon in these gravities can be cast as a first law of
thermodynamics which leads to suggest that the thermodynamic
interpretation of field equations at apparent horizon remains valid
even if one includes the possible quantum corrections to the
Einstein gravity. However, the thermodynamic interpretation of
gravity as a first law at the apparent horizon of FRW universe in
various gravity theories has been concentrated only for spherically
symmetric spacetime metric. If thermodynamic interpretation of
gravity near apparent horizon is to be generic, one needs to verify
whether the results may hold not only for more general spacetimes
but also for the other principles of thermodynamics, especially the
generalized second law as global accepted principle in the universe.
As a first step, we investigate further the thermodynamic of the
apparent horizon of FRW universe in the case of generalized second
law of thermodynamics. In this direction some work to suggest
constraints on the parameters of the gravity theory has been made
for the validity of second law \cite{wang}. A similar case for f(R)
gravity with equilibrium thermodynamics has been considered in
\cite{mohseni} and also see \cite{Barr,Gio,Zim} and reference
therein. By applying the general expression of temperature at
apparent horizon of FRW universe, it has been shown \cite{akbar}
that the generalized second law holds in Einstein gravity. Hence, it
is important to investigate whether the generalized law still holds
within a region of FRW universe enclosed by the apparent horizon
even if one includes the possible quantum corrections to the
Einstein gravity where the entropy-area relation no longer holds. In
this paper, We will discuss this issue in case of Gauss-Bonnet and
more general lovelock gravities. Let us now start with a
(n+1)-dimensional FRW universe of metric
\begin{equation}\label{1}
ds^{2}= -dt^{2}+
a^{2}(t)(\frac{dr^{2}}{1-kr^{2}}+r^{2}d\Omega^{2}_{n-1}),
\end{equation}
where $d\Omega^{2}_{n-1}$ stands for the line element of
(n-1)-dimensional unit sphere and the spatial curvature constant $k
= +1$, $0$ and $-1$ represents to a closed, flat and open universe,
respectively. The above metric (1) can be rewritten in spherical
form
\begin{equation}\label{2}
ds^{2}= h_{ab}dx^{a}dx^{b}+ \tilde{r}^{2}d\Omega^{2}_{n-1},
\end{equation}
where $\tilde{r} = a(t)r$, $x^{0} = t$, $x^{1}= r$ and $h_{ab} =
diag(-1,  \frac{a^{2}}{1-kr^{2}})$. One can work out the dynamical
apparent horizon from the relation
$h^{ab}\partial_{a}\tilde{r}\partial_{b}\tilde{r} = 0$ which turns
out
\begin{equation}
\frac{1}{\tilde{r}^{2}_{A}} = H^{2}+k/a^{2},
\end{equation}
where $H \equiv \frac{\dot{a}}{a}$ is the Hubble parameter and the
dots denote derivatives with respect to the cosmic time. It has been
found that the entropy $S$ associated with apparent horizon is
proportional to the horizon area that satisfies the so-called area
formula $S = A / 4G$ \cite{wal}, however it is well recognized that
the area formula of entropy no longer holds in higher derivative
gravities. So it is interesting to investigate whether, the
generalized second law, $\dot{S}_{h} + \dot{S}_{m} \geq  0$, still
hold or not for a region of a FRW universe enclosed by apparent
horizon if one consider higher derivative gravities, where $S_{h}$
is the entropy associated with the apparent horizon and $S_{m}$ is
the entropy of the source. Let us first consider Gauss-Bonnet
gravity which contains special combination of curvature-squared
term, added to the Einstein-Hilbert action. The equations of motion
for Gauss-Bonnet gravity are given by
\begin{equation}
G_{\mu\nu} + \alpha H_{\mu\nu} = 8\pi GT_{\mu\nu},
\end{equation}
where $\alpha$ is a constant with dimension $(length)^{2}$ and in
case of superstring theory of low energy limit, $\alpha$ is regarded
as the inverse string tension and is positive definite. $G_{\mu\nu}$
is the Einstein tensor, $T_{\mu\nu}$ is the stress-energy tensor of
perfect fluid and $H_{\mu\nu}$ is given by
\begin{equation}
H_{\mu\nu} = 2(RR_{\mu\nu} - 2R_{\mu\lambda}R^{\lambda}_{\nu} -
2R^{\gamma\delta}R_{\gamma\mu\delta\nu} +
R^{\alpha\gamma\delta}_{\mu}R_{\alpha\nu\gamma\delta} -
\frac{1}{2}g_{\mu\nu}R_{GB},
\end{equation}
where $G_{GB} = R^{2} - 4R_{\mu\nu}R^{\mu\nu} +
R_{\alpha\beta\mu\nu}R^{\alpha\beta\mu\nu}$ is the Gauss-Bonnet
term. In the vacuum Gauss-Bonnet gravity with / without a
cosmological constant, static black hole solutions have been found
and the thermodynamics associated with the black hole horizons have
been investigated in \cite{caij}. In this theory static spherically
symmetric black hole solution is given by
\begin{equation}
ds^{2} = -e^{\lambda (r)}dt^{2} + e^{\nu (r)}dr^{2} +
r^{2}d\Omega_{n-1}^{2},
\end{equation}
with
\begin{equation}
e^{\lambda (r)} = e^{-\nu (r)} = 1 +
\frac{r^{2}}{2\bar{\alpha}}\left(1 - \sqrt{1 + \frac{64\pi
G\bar{\alpha}M}{n(n-1)\Omega_{n}r^{n}}}\right),
\end{equation}
where $\bar{\alpha} = (n-2)(n-3)\alpha$ and $M$ is the mass of black
hole. As $\alpha \rightarrow 0$, the above metric leads to
Schwarzschild metric in Einstein gravity. The entropy \cite{caij}
associated with this black hole horizon is of the form, $S =
\frac{A}{4G}\left(1 +
\frac{(n-1)2\bar{\alpha}}{(n-3)r_{+}^{2}}\right)$, where $A =
n\Omega_{n}r_{+}^{n-1}$ is the horizon area and $r_{+}$ is the
horizon radius of the black hole. It has been argue in \cite{a9}
that the above entropy formula for black hole horizon also holds for
the apparent horizon of FRW universe in Gauss-Bonnet gravity but
replacing $r_{+}$ by the apparent horizon $\tilde{r}_{A}$ which
turns out
\begin{equation}
S_{h} = \frac{A}{4G}\left(1 +
\frac{(n-1)2\bar{\alpha}}{(n-3)\tilde{r}_{A}^{2}}\right),
\end{equation}
where $A = n\Omega_{n}\tilde{r}_{A}^{n-1}$ is the horizon area. The
Friedman equation in the Gauss-Bonnet gravity filled with perfect
fluid of stress energy ,$T _{\mu\nu} = (\rho + P)u_{\mu}u_{\nu} +
Pg_{\mu\nu}$, is given by
\begin{equation}
(H^{2} + \frac{k}{a^{2}}) + \bar{\alpha}(H^{2} +
\frac{k}{a^{2}})^{2} = \frac{16\pi G}{n(n-1)}\rho.
\end{equation}
In terms of the apparent horizon radius, the above equation yields
\begin{equation}
\frac{1}{\tilde{r}_{A}^{2}} +
\bar{\alpha}\frac{1}{\tilde{r}_{A}^{4}} = \frac{16\pi
G}{n(n-1)}\rho.
\end{equation}
Now, differentiating the above equation with respect to cosmic time
, one gets
\begin{equation}
\dot{\tilde{r}}_{A} = \frac{8\pi
G}{(n-1)}\frac{\tilde{r}_{A}^{3}H(\rho + P)}{(1 +
2\bar{\alpha}/\tilde{r}_{A}^{2})},
\end{equation}
here the continuity equation $\dot{\rho} = -nH(\rho + P)$ has been
used. It can be seen from the above equation that
$\dot{\tilde{r}}_{A} > 0$ provided dominant energy condition hold.
When $\alpha \rightarrow 0$, the above reduces to the result
obtained in Einstein gravity. Now we turn to define horizon
temperature which is proportional to surface gravity, $\kappa =
\frac{1}{2\sqrt{-h}}\partial_{a}(\sqrt{-h}h^{ab}\partial_{b}\tilde{r})$,
and is given by
\begin{equation}
T_{h}=|\kappa|/2\pi = \frac{1}{2\pi
\tilde{r}_{A}}(1-\frac{\dot{\tilde{r}}_{A}}{2H\tilde{r}_{A}}),
\end{equation}
where $\frac{\dot{\tilde{r}}_{A}}{2H\tilde{r}_{A}}\leq 1$ ascertains
that the associated temperature, $T_{h}$, is positive definite. Note
that the associated temperature, $T_{h}$, is independent of the
gravity theory but depends \cite{pwang} only on the background
geometry. Thus the above expression for temperature, $T_{h}$, also
holds in Gauss-Bonnet gravity. Let us now consider a region of a FRW
universe enveloped by the apparent horizon and assume that the
region bounded by the apparent horizon acts as a thermal system with
boundary defined by the apparent horizon and is filled up by a
perfect fluid of energy density $\rho$ and pressure $P$. Since the
apparent horizon is not constant but varies with time. As the
apparent radius changes , the volume enveloped by the apparent
horizon will also change, however the thermal system bounded by the
apparent horizon remains in equilibrium when it moves from one state
to another so that the temperature associate with the horizon must
be uniform and the same as the temperature of its surroundings. This
requires that the temperature of the source inside the apparent
horizon should be in equilibrium with the temperature associated
with the horizon. Hence we have $T_{m} = T_{h}$, where $T_{m}$ is
temperature of the energy enclosed by the apparent horizon. Let the
total energy of universe enclosed by the apparent horizon is the
total matter energy $E = V\rho$, where $V =
\Omega_{n}\tilde{r}_{A}^{n}$ is the volume enclosed by the apparent
horizon, where $\Omega_{n} = \pi^{n/2}/\Gamma(n/2 + 1)$ is the
volume of an n-dimensional unit ball. It has been shown that the
Friedmann equations of Gauss-Bonnet gravity can be recast \cite{a15}
as a thermal identity $T_{h}dS_{h} = dE + WdV$ at the apparent
horizon. Let us now turn to find out $T_{h}d{S}_{h} = \frac{1}{2\pi
\tilde{r}_{A}}(1-\frac{\dot{\tilde{r}}_{A}}{2H\tilde{r}_{A}})
d(\frac{n\Omega_{n}\tilde{r}_{A}^{n-1}}{4G}\left(1 +
\frac{(n-1)2\bar{\alpha}}{(n-3)\tilde{r}_{A}^{2}}\right))$ which on
simplifying yields
\begin{equation}
T_{h}\dot{S}_{h} = n\Omega_{n}\tilde{r}_{A}^{n}(\rho + P)H(1 -
\frac{\dot{\tilde{r}}_{A}}{2H\tilde{r}_{A}}).
\end{equation}
It is clear from equation (12) that the term, $(1 -
\frac{\dot{\tilde{r}}_{A}}{2H\tilde{r}_{A}}) > 0$, is positive to
ensure that the horizon temperature positive, however, in the
accelerated phase of the universe, $\rho + P$ may negative
indicating that the second law $\dot{S}_{h} > 0$ does not hold. But
we will show that in the case of generalized second law this is not
the case. It is well known that the Friedmann equations at apparent
horizon of FRW universe satisfied \cite{b1,a15,b2} the thermal
identity $dE = T_{h}dS_{h} + WdV$ instead of satisfying the standard
first law $T_{h}dS_{h} = dE + PdV$, where $T_{h}$ is the temperature
associated with the apparent horizon. However the entropy $S_{m}$ of
universe inside the apparent horizon can be connected to its energy
$E_{m} = V\rho = E$ and pressure $P$ at the horizon by the Gibbs
\cite{gibbs} equation, $d(V\rho) = T_{m}dS_{m} - PdV$, where $T_{m}$
is the temperature of the energy inside the horizon. we again assume
that the thermal system bounded by the apparent horizon remains in
equilibrium so that the temperature of the system must be uniform
and the same as the temperature of its surroundings. This requires
that the temperature $T_{m}$ of the energy inside the apparent
horizon should be in equilibrium with the temperature $T_{h}$
associated with the apparent horizon. Hence we have  $T_{m} = T_{h}
= |\kappa|/2\pi$ which is given by equation (12). Hence from the
Gibbs equation, $dE_{m} = T_{h}dS_{m} - PdV$, one can obtain
\begin{equation}
T_{h}\dot{S}_{m} = n\Omega_{n}\tilde{r}_{A}^{n-1}\dot{\tilde{r}}_{A}
- n\Omega_{n}\tilde{r}_{A}^{n}H(\rho + P)
\end{equation}
Adding equations (13) and (14), one gets
\begin{equation}
T_{h}(\dot{S}_{m} + \dot{S}_{h}) =
(\frac{n}{2}\Omega_{n}\tilde{r}_{A}^{n-1})(\rho +
P)\dot{\tilde{r}}_{A}.
\end{equation}
The right side of the above equation contains two factor. The first
factor $\frac{n}{2}\Omega_{n}\tilde{r}_{A}^{n-1} = \frac{1}{2}A$,
where $A$ is the horizon area, is positive definite. And the second
factor $(\rho + P)\dot{\tilde{r}}_{A} = \frac{8\pi
G}{(n-1)}\frac{\tilde{r}_{A}^{3}H(\rho + P)^{2}}{(1 +
2\bar{\alpha}/\tilde{r}_{A}^{2})}$ is also positive throughout the
history of the universe. Hence the above equation $(\dot{S}_{m} +
\dot{S}_{h}) \geq 0$ which implies that the generalized second law
holds for a Gauss-Bonnet gravity in a region enclosed by the
apparent horizon of FRW universe throughout the history of the
universe. When $\alpha = 0$, one gets that the generalized second
law also holds in Einstein gravity.\\
Let us now turn to derive the generalized second law in the case of
lovelock gravity. The lovelock gravity generalizes the Einstein
gravity when spacetime has a dimension greater than four. In this
case the most general lagrangian \cite{love} that gives second order
equations for the metric, is the sum over the dimensionally extended
Euler density $ \emph{L} = \sum_{n=0}^{m}c_{n}L_{n}$, where $c_{n}$
are arbitrary constants and $L_{n}$ is the Euler density of a
2n-dimensional manifold $L_{n} =
2^{-n}\delta^{\mu_{1}\nu_{1}....\mu_{n}\nu_{n}}_{\alpha_{1}\beta_{1}
...\alpha_{n}\beta_{n}}
R^{\alpha_{1}\beta_{1}}_{\mu_{1}\nu_{1}}...R^{\alpha_{n}\beta_{n}}_{\mu_{n}\nu_{n}}$,
where the generalized delta function
$\delta^{\mu_{1}\nu_{1}....\mu_{n}\nu_{n}}_{\alpha_{1}\beta_{1}
...\alpha_{n}\beta_{n}}$ is totally antisymmetric in both sets of
indices and $R^{\alpha\beta}_{\gamma\delta}$ are the components of
the curvature tensor. The static spherically symmetric black hole
solutions can be obtained \cite{wheeler} in this theory. For an
(n+1)-dimensional static spherically symmetric black hole of metric
\begin{equation}
ds^{2} = -f(r) dt^{2} + f^{-1}(r) dr^{2} + r^{2}d\Omega_{n-1}^{2},
\end{equation}
with metric function is given by $f(r) = 1-r^{2}F(r)$, where $F(r)$
is determined by solving for real roots of mth-order polynomial
equation, $\sum_{i = 0}^{m}\hat{c}_{i}F^{i}(r) = \frac{16\pi G
M}{n(n-1) \Omega_{n}}r^{n}$. Where $M$ is the mass of the black hole
and the coefficients $\hat{c}_{i}$ are $\hat{c}_{0} = \frac{c
_{0}}{n(n-1)}$, $c_{1} = 1$ and $\hat{c}_{i} =
c_{i}\prod_{j=3}^{2m}(n+1-j),~~~ for~~~i > 1$. The entropy
\cite{rcai} associated with the black hole horizon of lovelock
gravity can be expressed as $S = \frac{A}{4G}\sum_{i=1}^{m}
\frac{i(n-1)}{n-2i+1} \hat{c}_{i}r_{+}^{2-2i}$, where $A =
n\Omega_{n}r_{+}^{n-1}$ is the horizon area and $r_{+}$ is the
horizon radius. The above entropy formula of black hole horizon also
holds for the apparent horizon of FRW universe and the entropy
associated with the apparent horizon has the same expression but the
black hole horizon radius $r_{+}$ is replaced by the apparent radius
\cite{a9}. Hence the entropy expression for the apparent horizon
turns out
\begin{equation}
S = \frac{A}{4G}\sum_{i=1}^{m} \frac{i(n-1)}{n-2i+1}
\hat{c}_{i}\tilde{r}_{A}^{2-2i}.
\end{equation}
The Friedman equation \cite{a9} of a FRW universe in the Lovelock
theory of gravity is given by
\begin{equation}
\sum_{i=1}^{m}\hat{c}_{i}(H^{2} + \frac{k}{a^{2}})^{i} = \frac{16\pi
G}{n(n-1)}\rho.
\end{equation}
In terms of apparent horizon radius, $1/\tilde{r}_{A}^{2} = H^{2} +
k/a^{2}$, the above Friedman equation (18) can be expressed as
\begin{equation}
\sum_{i=1}^{m}\hat{c}_{i}(\tilde{r}_{A})^{-2i} = \frac{16\pi
G}{n(n-1)}\rho.
\end{equation}
From the above equation, one can get $\dot{\tilde{r}}_{A}$ by
differentiation it with respect to cosmic time which yields
\begin{equation}
\dot{\tilde{r}}_{A} = \frac{8\pi G}{(n-1)}\frac{H(\rho +
P)}{\sum_{i=1}^{m}i\hat{c}_{i}(\tilde{r}_{A})^{-2i-1}},
\end{equation}
which is positive provided the dominant energy condition holds.
Again we assume that the region of a FRW universe enclosed by the
apparent horizon acts as a thermal system with the apparent horizon
as a boundary of the system. In general the apparent radius is not
constant but change with time, however the system remains in
equilibrium when it moves from one state to another. This requires
that the temperature associated with the apparent horizon is in
equilibrium with the temperature of the energy enclosed by the
boundary of the system. Also the expression for the temperature,
$T_{h}=|\kappa|/2\pi = \frac{1}{2\pi
\tilde{r}_{A}}(1-\frac{\dot{\tilde{r}}_{A}}{2H\tilde{r}_{A}})$,
associated with the apparent horizon still holds in the lovelock
gravity because the temperature of associated with horizon depends
on the background geometry but it is independent of the gravity
theory. Let us now turn to find the expression for $T_{h}\dot{S}_{h}
= \frac{1}{2\pi
\tilde{r}_{A}}(1-\frac{\dot{\tilde{r}}_{A}}{2H\tilde{r}_{A}})
\frac{d}{dt}\left(\frac{n\Omega_{n}\tilde{r}_{A}^{n-1}}{4G}\sum_{i=1}^{m}
\frac{i(n-1)}{n-2i+1} \hat{c}_{i}\tilde{r}_{A}^{2-2i}\right)$ which
on simplifying yields
\begin{equation}
T_{h}\dot{S}_{h} = n\Omega_{n}\tilde{r}_{A}^{n}(\rho + P)H (1-
\frac{\dot{\tilde{r}}_{A}}{2H\tilde{r}_{A}}).
\end{equation}
The factor $(1- \frac{\dot{\tilde{r}}_{A}}{2H\tilde{r}_{A}})> 0$ of
the above equation is positive because of equation (12) and the
factor $n\Omega_{n}\tilde{r}_{A}^{n}(\rho + P)H$ is positive, zero
or negative depending on the phase of the universe which implies the
second law of thermodynamics does not hold in general. But for the
generalized second law this is not the case. It has been shown that
the entropy $S_{h}$ and the temperature $T_{h}$ associated with the
apparent horizon of FRW universe together with the Friedmann
equations satisfies \cite{b1,a15,b2} the thermal identity $dE =
T_{h}dS_{h} + WdV$ in Einstein, Gauss-Bonnet and Lovelock theories
of gravity. However the entropy $S_{m}$ of the universe inside the
apparent horizon can be connected to its energy $E_{m} = V\rho = E$
and pressure $P$ at the horizon by the Gibbs \cite{gibbs} equation,
$d(V\rho) = T_{m}dS_{m} - PdV$, where $T_{m}$ is the temperature of
the energy inside the horizon. We again assume that the thermal
system bounded by the apparent horizon remains in equilibrium so
that the temperature of the system must be uniform and the same as
the temperature of its surroundings. This requires that the
temperature $T_{m}$ of the energy inside the apparent horizon should
be in equilibrium with the temperature $T_{h}$ associated with the
apparent horizon. Hence we have  $T_{m} = T_{h} = |\kappa|/2\pi$
which is given by equation (12). Hence from the Gibbs equation,
$dE_{m} = T_{h}dS_{m} - PdV$, one can obtain
\begin{equation}
T_{h}\dot{S}_{m} = n\Omega_{n}\tilde{r}_{A}^{n-1}\dot{\tilde{r}}_{A}
- n\Omega_{n}\tilde{r}_{A}^{n}H(\rho + P)
\end{equation}
Adding equations (21) and (22), one gets
\begin{equation}
T_{h}(\dot{S}_{m} + \dot{S}_{h}) =
(\frac{n}{2}\Omega_{n}\tilde{r}_{A}^{n-1})(\rho +
P)\dot{\tilde{r}}_{A}.
\end{equation}
The right side of the above equation contains two factor. The first
factor $\frac{n}{2}\Omega_{n}\tilde{r}_{A}^{n-1} = \frac{1}{2}A$,
where $A$ is the horizon area, is positive definite. And the second
factor $(\rho + P)\dot{\tilde{r}}_{A} = \frac{8\pi
G}{(n-1)}\frac{H(\rho +
P)^{2}}{\sum_{i=1}^{m}i\hat{c}_{i}(\tilde{r}_{A})^{-2i-1}},$ is also
positive throughout the history of the universe. Hence the above
equation $(\dot{S}_{m} + \dot{S}_{h}) \geq 0$ which implies that the
generalized second law holds for a Lovelock gravity in a region
enclosed by the apparent horizon of FRW universe throughout the
history of the universe.
\paragraph{Conclusion:}
It has been shown that the Friedmann equations in Einstein,
Gauss-Bonnet and Lovelock gravities can be rewritten as a similar
form of the first law at the apparent horizon of FRW universe. Since
the higher derivative terms present in Gauss-Bonnet and Lovelock
gravities can be considered as a possible quantum corrections to the
Einstein gravity and the Friedmann equations at apparent horizon can
be interpreted as a first law in these gravities which leads to
suggest that the thermodynamic interpretation of gravity at apparent
horizon remains valid even one includes possible correction to the
Einstein gravity. To investigate further the thermodynamic
properties associated with the apparent horizon, one have to check
beyond the first law, especially the generalized second law as
accepted globally principe in the universe. For this purpose, it has
already been shown \cite{akbar} that the generalized second remains
valid in Einstein gravity at the apparent horizon of FRW universe.
However, in this paper, we extended this study for the Gauss-Bonnet
and Lovelock gravities and show that thermodynamic behavior of the
apparent horizon remains valid for the generalized second even one
includes possible quantum correction to the Einstein gravity. It is
shown that by employing the general expression of the temperature $T
= |\kappa| / 2\pi= \frac{-1}{2\pi
\tilde{r}_{A}}(1-\frac{\dot{\tilde{r}}_{A}}{2H\tilde{r}_{A}})$
associated with the apparent horizon of FRW universe and assuming
that the universe bounded by the apparent horizon remains in
equilibrium so that the temperature of the system must be uniform as
the temperature of its surroundings, the generalized second law
holds in Gauss-Bonnet gravity and more general Lovelock gravity in a
region enveloped by the apparent horizon. In this analysis, the
general expression of temperature, $T_{h}=|\kappa|/2\pi =
\frac{1}{2\pi
\tilde{r}_{A}}(1-\frac{\dot{\tilde{r}}_{A}}{2H\tilde{r}_{A}})$, has
been used to discuss generalized second law in various gravity
theories and no approximation for the horizon temperature has been
made. However, in the earlier work, to the best of my knowledge, an
approximate horizon temperature $T_{h} = 1 / 2\pi \tilde{r}_{A}$ has
been assumed to investigate the generalized second law (see for
examples \cite{Barr,Gio,Zim, mohseni,wang}). It is also important to
note that the horizon temperature $T_{h}$ is not uniform in general
to the temperature $T_{m}$ of the energy inside the horizon because
there may be flow of energy across the apparent horizon but in our
analysis, we are not considering the flow of energy across the
apparent horizon. However, some authors (see for example
\cite{davies}) have assumed $T_{m} = bT_{h}$, where $b > 0$ to
assure that the temperature of the system is positive. It is also
important to investigate the generalized second law further for f(R)
and scalar-tensor gravities at the apparent horizon of FRW universe.
\section*{Acknowledgments} I would like to thank Rong-Gen Cai for
his useful comments. The work is supported by a grant from National
university of Sciences and Technology.


\end{document}